\documentclass[aps,pra,preprint,groupedaddress]{revtex4-1}
\usepackage[latin1]{inputenc}
\usepackage{amsmath}
\usepackage{amsfonts}
\usepackage{amssymb}
\usepackage{graphicx}

\begin{document}

\title{Evaluation of the phase randomness of the light source in quantum key distribution systems with an attenuated laser}

\author{Toshiya Kobayashi}
\altaffiliation[Present address: ]{Seiko Epson Corporation}
\author{Akihisa Tomita}
\author{Atsushi Okamoto }
\affiliation{Graduate School of Information Science and Technology, Hokkaido University\\
Kita 14, Nishi 9, Sapporo 060-0814, Japan}

\begin{abstract}
The phase randomized light is one of the key assumptions in the security proof 
of Bennett-Brassard 1984 (BB84) quantum key distribution (QKD) protocol 
implemented with an attenuated laser. 
Though the assumption has been believed to be satisfied for conventional systems,
 it should be reexamined for current high speed QKD systems.
The phase correlation may be induced by the overlap of the optical  pulses, 
 the interval of which decreases as the clock frequency.
The phase randomness was investigated experimentally 
 by measuring the visibility of interference.
An asymmetric Mach-Zehnder interferometer was used to observe the interference 
 between adjacent pulses from a gain-switched distributed feedback laser diode 
 driven at 10 GHz.
Low visibility was observed when the minimum drive current was set far below
 the threshold, while the interference emerged when the minimum drive current
  was close to the threshold.
Theoretical evaluation on the impact of the imperfect phase randomization
 provides target values for the visibility to guarantee the phase randomness.
The experimental and theoretical results show that secure implementation of
 decoy BB84 protocol is achievable even for the 10-GHz clock frequency,
  by using the laser diode under proper operating conditions.
\end{abstract}

\maketitle

\section{Introduction}
Quantum key distribution (QKD) offers an unconditionally secure method
 to share a cryptographic key between remote parties. 
Bennett-Brassard 1984 (BB84) protocol \cite{BB84} is one of the most developed
 QKD protocols, the security proof of which has been well established \cite{Shor2000spo,Koashi2005,Renner2008soq,Koashi2009ssp}.  
Recent researches on the security focus on more practical aspects,
 such as imperfections in a QKD apparatus.  
In actual QKD equipment, the device characteristics deviate from the ideal ones. 
Keeping the secure key rate with imperfect devices is an important issue
 \cite{Koashi2003sqk, Scarani2009sop}.
Since a practical single photon source, assumed in the original BB84 protocol,
 is still unavailable, most QKD experiments have utilized light pulses
  from a laser diode (LD) after strong attenuation. 
The attenuated laser pulses contain two photons or more with a finite probability. 
The multiple photon states opened the way to an efficient eavesdropping method
 called photon number splitting (PNS) attack \cite{Brassard2000lpq}. 
Gottesman, Lo, L\"{u}tkenhaus, and Preskill (GLLP \cite{GLLP2004}) analyzed
 the security against this imperfection.
An improved protocol called decoy-BB84  \cite{Hwang2003qkd,Wang2005bpa,Lo2005dsq}
 was proposed to yield better secure key rate than GLLP. 
The decoy-BB84 protocol provides not only unconditional security
 with the attenuated laser light, but also the universal composability
 \cite{ben2005universal,Muller-Quade2009ciq,hayashi2012con,Tomamichel2012tfa}.

The strongly attenuated laser light is often called weak coherent light. 
This term is misleading, because the security analysis in the GLLP
 and decoy-BB84 articles assumes that the light source emits  photons
 in a  phase randomized Poissonian state,
 which is a mixture of coherent states with uniformly distributed phases:
\begin{equation}
 \rho=\frac{1}{2 \pi} \int_{-\pi}^{\pi} d \phi
  |\alpha e^{i \phi} \rangle \langle \alpha e^{i \phi}| 
 = e^{-|\alpha|^2} \sum_{n=0} ^{\infty} \frac{\alpha^{2n}}{n!}|n
  \rangle \langle n|. 
 \label{Poisson}
\end{equation}
The state is represented by a diagonal density matrix with respect to
 photon-number basis.  
Lo and Preskill \cite{Lo2007soq} showed that, if the photon states were
 really weak coherent,
  discrimination of the bases used in the BB84 protocol would be easier.
Recently, Tang, et al. \cite{Tang2013sao} showed that
 the phase information also increases distinguishability
  between decoy and signal pulses used in the decoy-BB84 protocol. 
Those reports have issued a warning about the phase correlation
 among the laser pulses;
 the phase correlation will increase the information leakage
  and thus reduce the secure key rate. 
Active phase randomization was proposed and implemented
 by Zhao, et al. \cite{Zhao2007eqk}.
Effect of partially randomized phase was also examined
 for a plug-and-play system \cite{Sun2012prp}.

Nevertheless, most experimentalists have not taken this warning seriously
 with a few exceptions \cite{Tang2013sao, Zhao2007eqk, Sun2012prp}.
Their common belief is that  pulses from a gain-switched LD have
 no phase relationship to other pulses.
Therefore, the phase of the light source is automatically randomized,
 as long as the one-way QKD architecture is employed.
The mechanism of the phase randomization is following.   
In the gain-switched mode, each current pulse excites the semiconductor medium
 from loss to gain. 
A laser pulse is generated from seed photons originated
 from spontaneous emission,
 because the photons from the previous lasing have vanished
 during the pulse interval. 
The phase of the spontaneous emission is random,
 so that the phase of the laser pulses should vary
  from one pulse to another.  
This is true when the previously lased photons disappear completely
 in the interval.  
However, if the  photons survive until the next  excitation,
 the lasing can be seeded by the remaining photons.  
Then, the phase of the laser pulse may relate to the previous one,
 because the stimulated emission conserves the phase.  
The effect of the residual photons will emerge significantly by increasing
 the pulse repetition rate and narrowing the pulse interval.
The state-of-art  QKD systems operate at high clock frequencies over 1 GHz,
 along with the improvement of the photon detectors
  \cite{Dixon2008gdq,Sasaki2011fto,Yoshino2012hwm}.  
The clock frequency would further increase to meet demands
 for high bit-rate secure communication. 
The interval time thus decreases down to hundreds picoseconds or even shorter. 
Furthermore, the drive current may not return to zero
 in order to improve modulation response of the laser.
It is unclear whether the assumption of the phase randomized source still holds
 in QKD systems operated at several GHz-clock frequencies. 
 
In this article, we examine the phase randomness of the light source
 at 10-GHz clock frequency.
Sec. \ref{sec:theory_of_measurement} introduces an asymmetric interferometer
 set-up to measure the phase correlation between the adjacent optical pulses. 
We recall the relation of the phase correlation to the visibility of
 the interference fringe. 
Sec. \ref{sec:impact} considers the effects of the partial coherence
 in state discrimination, 
 which were analyzed by Lo-Preskill \cite{Lo2007soq}
 and Tang et al. \cite{Tang2013sao} for perfectly coherent states.
We provide target values of the visibility, under which we can regard
 the light source as phase randomized.
Sec. \ref{sec:exp} shows the  measured visibility of the interference fringe of
 the adjacent pulses from a LD operated at 10-GHz clock frequency. 
We controlled DC bias current to the LD,
 which determines the effective pulse interval and the minimum drive current.
In sec. \ref{sec:discussion}, we examine the accuracy of the estimated values
 of visibility, and applied corrections to the estimation.
We investigate the relation between the observed phase correlation
 and the operating conditions,
 in terms of the effective photon life time of the LDs.   

\section{Theory} \label{sec:theory}
\subsection{Relation between visibility and phase correlation}
\label{sec:theory_of_measurement}
\begin{figure}
\includegraphics[width=0.8\linewidth]{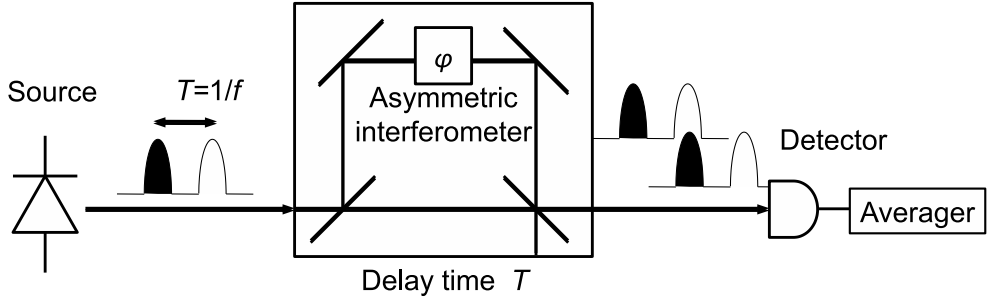}
\caption{A schematic illustration of a measurement apparatus
 for the phase correlation between the adjacent pulses. 
The delay time of the asymmetric interferometer is adjusted to the pulse period. 
The phase difference $\varphi$ between the paths can be modulated.}
\label{fig:schematics}
\end{figure}

The phase relation of laser light can be characterized with an interferometer. 
Figure \ref{fig:schematics} illustrates a schematic of
 an asymmetric interferometer to observe interference
  between the adjacent optical pulses.
We focus on measuring the interference between the adjacent pulses,
 because the phases between the adjacent pulses are more correlated than
  those between more temporally-separated pulses. 
Light pulses generated in the source enter the asymmetric interferometer,
 where the delay time is adjusted to the pulse period. 
The adjacent optical pulses are combined at the output. 
A phase modulator is placed in one arm of the interferometer
 to provide a phase difference $\varphi$ between the paths.
The signals are detected by a high-speed photodetector and
 accumulated by an averager. 
If a fixed phase relation  between the adjacent pulses exists, 
 the amplitude of the signal takes a definite value according
  to the phase difference between the optical paths.
A clear interference fringe will be observed as $\varphi$ varies.
If the phases between the pulses are random, the interference signal differs
 from pulse to pulse. 
Then the interference fringe will disappear after accumulation. 
Visibility of interference $\Theta$,
 which represents the degree of the phase correlation, is defined by
\begin{equation}
0 \leq \Theta := \frac{I_{max}-I_{min}}{I_{max}+I_{min}} \leq 1,
\label{visibility}
\end{equation}
where $I_{max}$ and  $I_{min}$  stand for the peak (maximum) and
  the valley (minimum) of the interference fringe, respectively. 
As the phase correlation becomes stronger, the visibility gets closer to one.

In the following, we recall a relation between the visibility
 and the phase correlation \cite{YarivYeh2006}.
The output intensity of the interferometer is given by  
\begin{equation}
I(\varphi)\propto \mathcal{E}^2 \left\{ |a_{A}|^2+|a_{B}|^2+|a_{A}a_{B}|
 \exp \left[i (\theta + \varphi) \right]+\mathrm{c.c.} \right\},
\label{I(x)}
\end{equation}
where the coefficient $\mathcal{E}=\sqrt{\hbar\omega/2\epsilon_0V}$
 carries the dimension of electric field.
The complex amplitudes $a_A$ and $a_B$ represent the fields provided
 from paths $A$ and $B$, respectively. 
In the asymmetric interferometer, the fields  $a_A$ and $a_B$ correspond
 to those of the adjacent pulses.
The relative phase between the pulses is given by $\theta$.
The third and forth terms of Eq. (\ref{I(x)}) are responsible
 for the interference. 
Taking an average over an ensemble, we obtain the interference terms as
\begin{equation}
\left\langle a_{A}a_{B}\right\rangle \left\langle e^{i\theta} \right\rangle
 e^{i\varphi}  +\mathrm{c.c.},
\label{jikanheikin}
\end{equation}
where we assume the phase difference $\varphi$ varies slowly,
 while the relative phase $\theta$ is a probabilistic variable. 
Equation (\ref{jikanheikin}) shows that the interference visibility is governed
 by the expectation value of the relative phase $\langle e^{i\theta} \rangle$.
If the distribution of the phase obeys a Gaussian probability density function
 with the central value $\theta_0$ and the standard deviation $\sigma$
\begin{equation}
g(\theta, \theta_0)=\frac{1}{\sqrt{2\pi\sigma^2}}
 \exp \left[-\frac{(\theta-\theta_0)^2}{2\sigma^2}\right],
\end{equation}
the expectation value is given by 
\begin{align}
\left\langle e^{i\theta}\right\rangle
 &=\int_{-\infty}^{\infty}e^{i\theta}g(\theta,\theta_0)d(\theta) \nonumber \\
 &= \exp\left[-\frac{\sigma^2}{2} + i \theta_0 \right].
\label{kitaichi}
\end{align}
The Gaussian probability distribution describes the phase distribution of
 the  LD light well,
 because the light field vector  (phaser) in the phase space is kicked
  by a number of photons generated by spontaneous emission \cite{Henry1982tol}. 
The kicks force the field vector to walk randomly around the original position. 
Since the spontaneous emission occurs independently,
 a number of kicks results in the Gaussian distribution.
Using Eqs. (\ref{visibility})-(\ref{kitaichi}), we relate the visibility
 to the standard deviation of the phase distribution as
\begin{equation}
\Theta =  \exp \left[-\frac{\sigma^2}{2} \right]. 
\label{eq:visigma}
\end{equation}
Here, we assume $|a_{A}|=|a_{B}|$ for simplicity. 
As expected, the visibility decreases rapidly
 with increasing the standard deviation of the phase.
The analysis given above uses the classical complex amplitude
 of the electric field, 
 because the laser field can be well approximated with a classical field,
 where the effect of the spontaneous emission is introduced by a random kick
 \cite{YarivYeh2006,Henry1982tol}. 
The wave properties of the field will not be altered by attenuation.
Furthermore, the analysis using quantum operators will provide
 the similar description. 
  
\subsection{Impact of phase correlation} 
\label{sec:impact}   
We consider the impact of phase correlation
 to provide criteria to guarantee the security of decoy-BB84
  with a LD light source. 
As stated in the introduction, the phase correlation enhances
 the distinguishability of the states,
 which are expected to be indistinguishable in the ideal situation
  \cite{Lo2007soq, Tang2013sao}. 
The following calculation will treat two issues on the state discrimination:
 one is between the states of different bases,
  and the other is between the signal and decoy pulses.

Most security proofs of BB84 rely on the assumption that the density matrix of
 one basis is indistinguishable to another. 
The distingushability of two density matrices, sometimes called the imbalance of
 the quantum coin \cite{GLLP2004}, 
 helps the eavesdropper (Eve) to distinguish the state encoding. 
GLLP \cite{GLLP2004} described the imbalance in terms of the fidelity
 between the density matrices, and analyzed its effects on the security. 
Though the imbalance of the quantum coin often refers to
 the state preparation flaws, Lo and Preskill \cite{Lo2007soq} showed
  that phase correlation also enhances the distingushability. 
The imbalance of the quantum coin $\Delta$ is given by
\begin{equation}
 \Delta=\frac{1-F(\rho_X,\rho_Z)}{2},
 \label{imbalance}
\end{equation} 
where the fidelity of the density matrices in $X$-coding and $Z$-coding
 is defined by
\begin{equation}
 F(\rho_X,\rho_Z)=\mathrm{Tr} \left( \rho_Z^{1/2} \rho_X \rho_Z^{1/2}\right)^{1/2}.
 \label{fidelity}
\end{equation}
Further, since Eve may exploit the channel loss, 
 we should recalculate the imbalance to keep security as
\begin{equation}
 \Delta^{\prime} =\frac{\Delta}{\eta \mu}
 \label{imbalance2}
\end{equation}
 for given transmittance of the channel $\eta$
  and the average photon number $\mu$ of the source. 
Recently, Tamaki, et al. \cite{Tamaki2013lqc} showed that GLLP analysis
 was too conservative, and proposed a loss-tolerant proof
  even with state preparation flaws. 
We here calculate $\Delta$, because it still provides
 a comprehensive measure of the state distinguishability.
We can obtain $\Delta^{\prime}$ from $\Delta$ by Eq. (\ref{imbalance2}),
 if we follow the GLLP analysis. 
The density matrices of the partially phase randomized coherent states
 are expressed by
\begin{align}
\rho_Z =& \frac{1}{2} \int 
\left( \left|\sqrt{\mu} e^{i \theta} \right \rangle_F \left\langle \sqrt{\mu} e^{i \theta}\right|
\otimes \left|0 \right\rangle_S \left\langle 0 \right| + \left|0 \right\rangle_F \left\langle 0 \right| \otimes \left|\sqrt{\mu} e^{i \theta} \right\rangle_S \left\langle \sqrt{\mu} e^{i \theta}\right| \right)
g(\theta,\theta_0) d \theta \nonumber \\
=& \frac{1}{2} e^{-\mu}\left( \sum_{M,N}\frac{\mu^{(M+N)/2} e^{-(M-N)^2 \sigma^2/2}e^{i(M-N)\theta_0}}{\sqrt{M! N!}} |M \rangle_F \langle N|\otimes |0 \rangle_S \langle 0| \right.\nonumber \\
&+ \left. |0 \rangle_F \langle 0| \otimes \sum_{M,N}\frac{\mu^{(M+N)/2} e^{-(M-N)^2 \sigma^2/2}e^{i(M-N)\theta_0}}{\sqrt{M! N!}} |M \rangle_S \langle N|\right) \label{rhoZ}\\
\rho_X =& \frac{1}{2} \int \left( \left|\sqrt{\frac{\mu}{2}} e^{i \theta} \right\rangle_F \left\langle \sqrt{\frac{\mu}{2}} e^{i \theta}\right|
\otimes \left|\sqrt{\frac{\mu}{2}} e^{i \theta} \right\rangle_S \left\langle \sqrt{\frac{\mu}{2}} e^{i \theta}\right|  \right. \nonumber \\
&+ \left.\left|\sqrt{\frac{\mu}{2}} e^{i \theta} \right\rangle_F \left\langle \sqrt{\frac{\mu}{2}} e^{i \theta}\right| \otimes \left|-\sqrt{\frac{\mu}{2}} e^{i \theta} \right\rangle_S \left\langle - \sqrt{\frac{\mu}{2}} e^{i \theta}\right| \right) g(\theta,0) d \theta \nonumber \\
=& \frac{1}{2} e^{-\mu} \sum_{M,N}\left(\frac{\mu}{2}\right)^{(M+N)/2} e^{-(M-N)^2\sigma^2/2} \nonumber \\
&\times \sum_{m,n}\frac{1+(-1)^{m-n}}{\sqrt{(M-m)!m!(N-n)!n!}}|M-m \rangle_F\langle N-n | \otimes |m \rangle_S \langle n|, \label{rhoX}
\end{align}    
where the subscripts $F$ and $S$ denote fast and slow components of
 the time-bin qubits. 
A finite value of $\theta_0$ is assumed for $Z$ coding,
 while it is set to zero for $X$ coding.
Since only the relative phase between the two coding affects
 the distinguishability, this setting will not lose generality.
The factor $\exp[-(M-N)^2\sigma^2/2]$ decreases rapidly for large $\sigma$,
 and only the $M=N$ terms survive.  
In this phase randomized limit, the density matrices (\ref{rhoZ})
 and (\ref{rhoX}) coincide with those of the mixture of $M$-photon number states
  after the state preparation.

\begin{figure}
\centering
\includegraphics[width=0.8\linewidth]{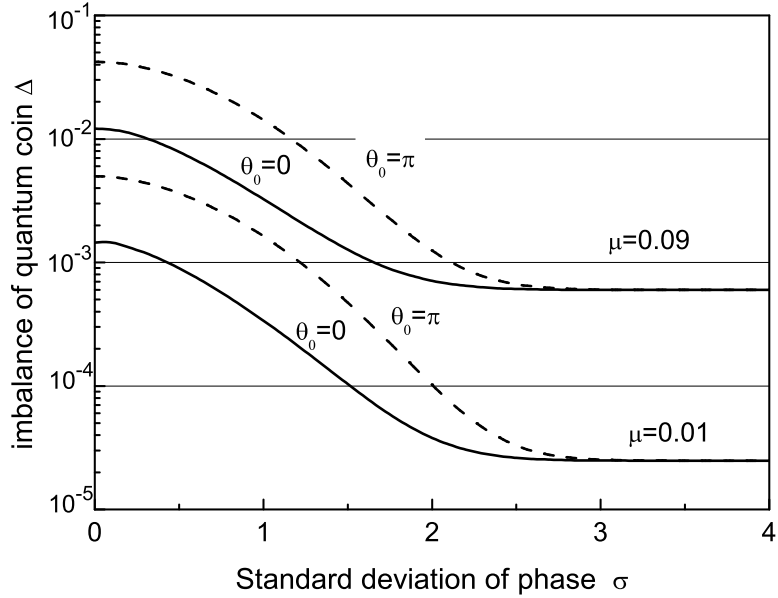}
\caption{The imbalance of the quantum coin,
 defined by $\Delta=(1-F(\rho_Z,\rho_X))/2$,
  as a function of standard deviation of phase distribution.
  Solid lines represent $\Delta$ for the difference of central phase value
   $\theta_0=0$, broken lines for $\theta_0=\pi$.}
\label{fig:imbalance}
\end{figure}

The imbalance of the quantum coin $\Delta$ can be calculated
 with Eq. (\ref{imbalance})
  and the density matrices (\ref{rhoZ}) and (\ref{rhoX}). 
The two states $\rho_Z$ and $\rho_X$ are most distinguishable
 when the central value of the phase $\theta_0=\pi$,
  and least distinguishable when $\theta_0=0$. 
We calculated the fidelity numerically with the density matrices
 in the photon-number-state basis truncated to a finite photon number $N_{max}$. 
We changed the number of bases to check the accuracy of the calculation. 
Since the average photon number is small, the results converged rapidly
 at $N_{max}=8$. 
We thus set  $N_{max}=16$ in the following calculation.
Figure \ref{fig:imbalance} shows the calculated values of $\Delta$
 as a function of the standard deviation of the phase distribution $\sigma$
  for the average photon numbers $\mu$ of 0.09 and 0.01. 
The imbalance of the quantum coin decreases as the standard deviation. 
For a small standard deviation, that is, less phase randomized, the imbalance of
 the quantum coin for $\theta_0 = \pi$  is larger than that for  $\theta_0 = 0$.  
The effect of  the central phase difference vanishes
 for large standard deviation, as the phases of the states become randomized.  
As seen in Fig.  \ref{fig:imbalance}, $\Delta$ converges to finite values
 for large standard deviation.
The relative errors from the asymptotic values fall below $10^{-2}$,
 when the standard deviation exceeds the following values:
 2.9 for $\mu=0.01$ and $\theta_0 = 0$, 2.6 for  $\mu=0.01$ and $\theta_0 =\pi$,
  3.2 for  $\mu=0.09$ and $\theta_0 = 0$,
    and 2.9 for $\mu=0.09$ and $\theta_0 = \pi$.
The values of standard deviation, 2.6, 2.9, and 3.2, 
 correspond to the visibility of 0.034, 0.015, and 0.006, respectively,
  which are estimated with Eq. (\ref{eq:visigma}). 
Therefore, target visibility values would be 0.015 for $\mu=0.01$ and 0.006
 for $\mu=0.09$ in terms of the imbalance of the quantum coin. 
The asymptotic value of  $\Delta$  for $\mu=0.09$ is larger than that
 for $\mu=0.01$, due to the  multiple photon contribution,
  which increases the distinguishability between the two states.  

Decoy method uses the states with different average photon numbers,
 called  signal and decoy. 
A key assumption of the decoy method is that Eve cannot distinguish
 the signal pulses from the decoy.
In other words, Eve can measure the photon number contained in a pulse,
 but cannot measure the average photon number of an individual pulse. 
Then Eve's strategy is limited to the one that depends 
 on the photon number of the pulse. 
The security proof of the decoy method only needs to consider such limited
 eavesdropping strategy.
Tang, et al. \cite{Tang2013sao} pointed out that phase correlation enables
  an unambiguous-state-discrimination (USD) measurement to distinguish 
   the signal from the decoy. 
The final key generated by the non-phase-randomized system can be compromised
 by combing the USD measurement and the PNS attack.  
When the phase is partially randomized, the USD measurement is
 no longer possible. 
However, if Eve allows finite probability to obtain inconclusive
 results $P_{inc}$,
 she can still increase the probability of the correct decision $P_C$
  \cite{FJ2003odo}. 
In the appendix, we derive the optimum positive-operator valued measure (POVM)
 to discriminate the signal state from the decoy state
  for partially coherent states.
The results of the POVM enable to extend the analysis presented by Tang,
 et al. \cite{Tang2013sao}.  

\begin{figure}
\centering
\includegraphics[width=0.8\linewidth]{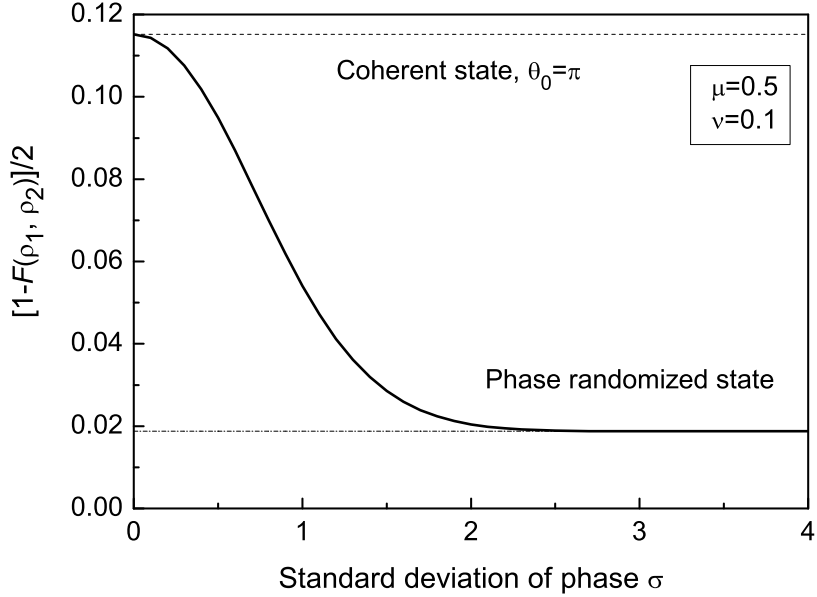}
\caption{Distinguishability $(1-F(\rho_1,\rho_2))/2$ between the signal and
 decoy states as a function of the standard deviation of phase. 
Calculation was done for the partially phase randomized states of
 average photon number 0.5 (signal) and 0.1 (decoy.)
Solid line represents the calculated fidelity for $\theta_0=\pi$.
Dash-dot line stands for the distinguishability
 between the completely phase randomized states, and dash line
  between coherent states with  $\theta_0=\pi$. }
\label{fig:fid_decoy}
\end{figure}

We investigate the fidelity between the signal and decoy states described by 
\begin{align}
 \rho_1 &= e^{-a_1^2} \sum_{m,n}\frac{a_1^{m+n} e^{-(m-n)^2 \sigma^2/2}
  e^{i(m-n)\theta_0}}{\sqrt{m! n!}} |m \rangle \langle n| \\
 \rho_2 &= e^{-a_2^2} \sum_{m,n}\frac{a_2^{m+n} e^{-(m-n)^2 \sigma^2/2}}{\sqrt{m! n!}} |m \rangle \langle n|.
 \label{eq:decoy}
\end{align}
Following Tang, et al. \cite{Tang2013sao},
  we consider only the fast component of the time-bin qubits,
   which carries no information on the key bit value. 
The density matrices $\rho_1$ and $\rho_2$ describe
 partially phase randomized coherent states
 with the average photon numbers $\mu=2 a_1^2$ and $\nu=2 a_2^2$, respectively. 
We calculated  the fidelity numerically by truncating the number of basis
 to a finite photon number $N_{max}=16$, as the imbalance of the quantum coin. 
Figure \ref{fig:fid_decoy} shows the distinguishability defined
  by $(1-F(\rho_1, \rho_2))/2$ with the fidelity of $\rho_1$ and $\rho_2$. 
The distinguishability decreases as the phase randomization,
and asymptotically reaches the value for the completely phase randomized states.
The relative discrepancy between the two became less than
 $10^{-2}$ for $\sigma > 2.5$, which corresponds to the visibility of 0.044.
This value would be a target visibility in terms of the signal-decoy
 discrimination.
The fidelity was calculated for $\theta_0 = \pi$, because the coherent states
 with the relative phase $\theta_0 = 0$  yield the same fidelity
  as the completely phase randomized states. 
However, the coherent states still  provide an advantage to the eavesdropper
 to perform an individual attack as seen in the appendix.
 
 In this section, we have derived criteria of the phase randomness
  and thus the interference visibility.
 The target values  depend on the average photon numbers. 
 Moreover, the eavesdropping methods  are not exhausted
  with those considered above, so that the target values may be further lowered.
 Nevertheless, we believe that the present analysis covers a wide range of
  eavesdropping, and the values estimated here should be good indications.  
 
\section{Experiment}
\label{sec:exp}

The phase correlation measurement system consists of the interferometer and
 the pulsed light source to be tested.
We employed the configuration similar to the one depicted
 in Fig. \ref{fig:schematics}.
The source was a distributed feedback (DFB) LD (NEL, NLK5C5EBKA,)
 which was designed for 10-GHz direct modulation to emit optical pulses
  in a single longitudinal and transversal mode. 
It lases around the wavelength of 1560 nm at the threshold current of 9.5 mA.
The LD was driven by the combination of a 10-GHz sinusoidal current ($I_{AC}$)
 and a DC bias current ($I_{DC}$.)
The total current to the LD is expressed by $I_{AC}+I_{DC}$. 
The sinusoidal current injected to the laser is expressed by
\begin{equation}
 I_{AC} = \frac{I_{pp} }{2} \cos (2 \pi f t +\phi_{LD}),
 \label{current}
\end{equation}
where $I_{pp}$ stands for the peak-to-peak value of the sinusoidal current. 
The current changes periodically with the frequency  $f=10$ GHz
 and an initial phase $\phi_{LD}$.
The sinusoidal signal from a pulse-pattern-generator (PPG) was amplified
 to a fixed amplitude $V_{pp}$=4.615 V.
We estimated the peak-to-peak AC current to the LD as $I_{pp}$=92.3 mA,
 considering the 50-$\mathrm{\Omega}$ road resistance.
However, in high frequency region such as 10 GHz,
 the emerging effects of parasitic impedances of the LD and the circuit
  may reduce the current injected into the LD active layer. 
To correct this effect, we measured the modulation response of the LD
 with a network analyzer and a 45-GHz band-width photodetector.
The resonant-like frequency was about 12 GHz in this measurement,
 so that the intrinsic response of the LD affects little
  the modulation response up to 10 GHz.
It was found that the optical power response of the LD  decreased by about 1 dB
 at 10 GHz from that at 100 MHz. 
Since the optical power is proportional to the  injected current,
 we regard  the reduction of the response as the decrease of the current
  with the same  proportion. 
Then the net current $I_{net}$ is reduced from the nominal value
 $I_{nom}$ by $10 \log_{10} \left(I_{net}/I_{nom}\right)=-1$. 
The net AC current thus swung by $I_{pp}=92.3 \times  0.794= 73.3$ mA,
 where  $10^{-1/10} \simeq 0.794$.

The operating condition of the LD was controlled by changing
 the DC bias current $I_{DC}$.
We define the minimum drive current defined by $I_{min}=-I_{pp}/2 +I_{DC}$,
 which refers to the drive current at the bottom of the AC current. 
In the following, we use normalized excitation
 to describe the operating condition.
The normalized minimum excitation is defined by
\begin{equation}
\Lambda = \frac{I_{min}-I_{th}}{I_{th}}. 
\end{equation}
As mentioned, the laser threshold current was $I_{th}=$ 9.5 mA. 
When $\Lambda>0$, the LD was always turned on.  
When $\Lambda<0$, the LD was turned off during the pulse interval.
The turn-off duration increases as the DC bias current decreases,
 which is obtained as a solution of $I_{AC}+I_{DC}=I_{th}$
  with Eq. (\ref{current}).
When $\Lambda<-1$, the LD was reversely biased and no current was injected at the minimum. 
In the present experiment, $\Lambda=0$ and $\Lambda=-1$ correspond to
 $I_{DC}$=46.15 mA and $I_{DC}$=36.75 mA, respectively.

\begin{figure*}
\centering
\includegraphics[width=0.8\linewidth]{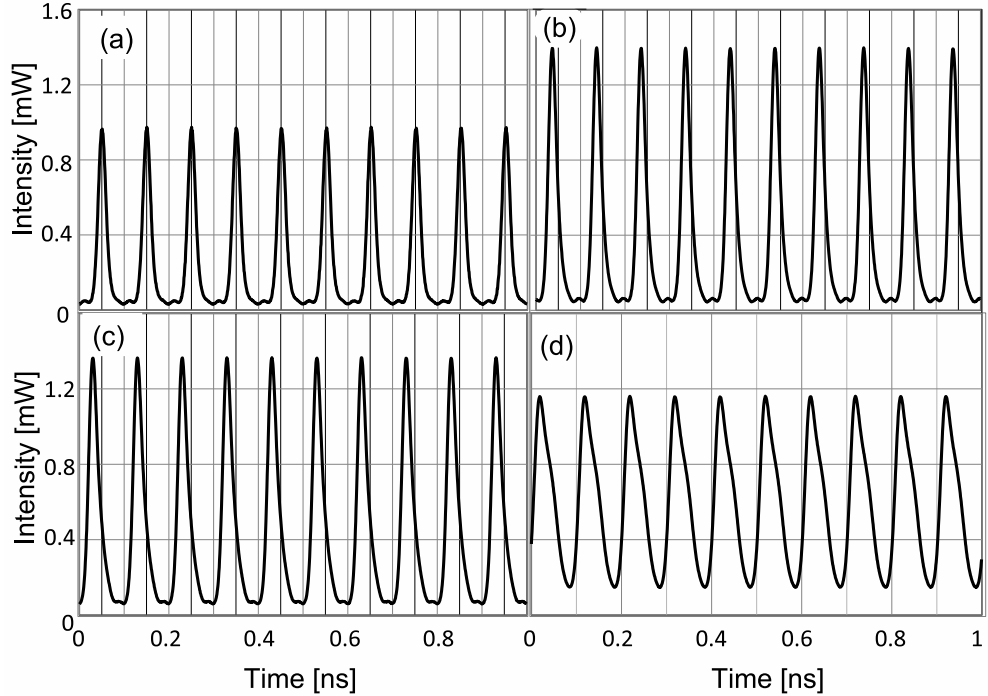}
\caption{Waveforms of the LD pulses. 
 The values of the normalized minimum excitation are as follows:
  (a) $\Lambda=-1.6$,  (b) $\Lambda=0.074$, (c) $\Lambda=0.49$,
   and (d) $\Lambda=2.6$.}
\label{fig:lightpulse}
\end{figure*}

We employed a commercially available asymmetric Mach-Zehnder interferometer
 (AMZI) module (Kylia, WT-MINT-M-L) to obtain interference
  between the adjacent pulses at 10 GHz, which was developed as a demodulator
   for 10-GHz differential phase shift keying (DPSK.)
The phase difference between the optical paths was modulated
 with a phase shifter integrated in the AMZI module.
The signal was accumulated for 256 samples and measured with a sampling
 oscilloscope of 40-GHz optical band-width to observe the interference fringe.
The output of the AMZI was attenuated by an optical attenuator
 to avoid saturation of the photodetector.
The peak and valley intensities of the accumulated interference fringes
 were recorded. 
 
Figure \ref{fig:lightpulse} shows the observed waveforms of light pulses
 for (a) $\Lambda=-1.6$,  (b) $\Lambda=0.074$, (c) $\Lambda=0.49$,
  and (d) $\Lambda=2.6$, where the minimum drive current $I_{min}$ was
   (a) below the threshold, (b) near the threshold, (c) above the threshold,
    and (d) far above the threshold. 
By setting the $I_{min}$ close to the threshold, sharp and intense pulses were
 obtained as shown in Fig. \ref{fig:lightpulse} (b) and (c). 
When the $I_{min}$ was far above the threshold, the laser output reflected
 the input current waveform as in Fig. \ref{fig:lightpulse} (d). 
The LD was no longer operated in the gain-switched mode in this DC bias region. 

\begin{figure*}
\centering
\includegraphics[width=0.8\linewidth]{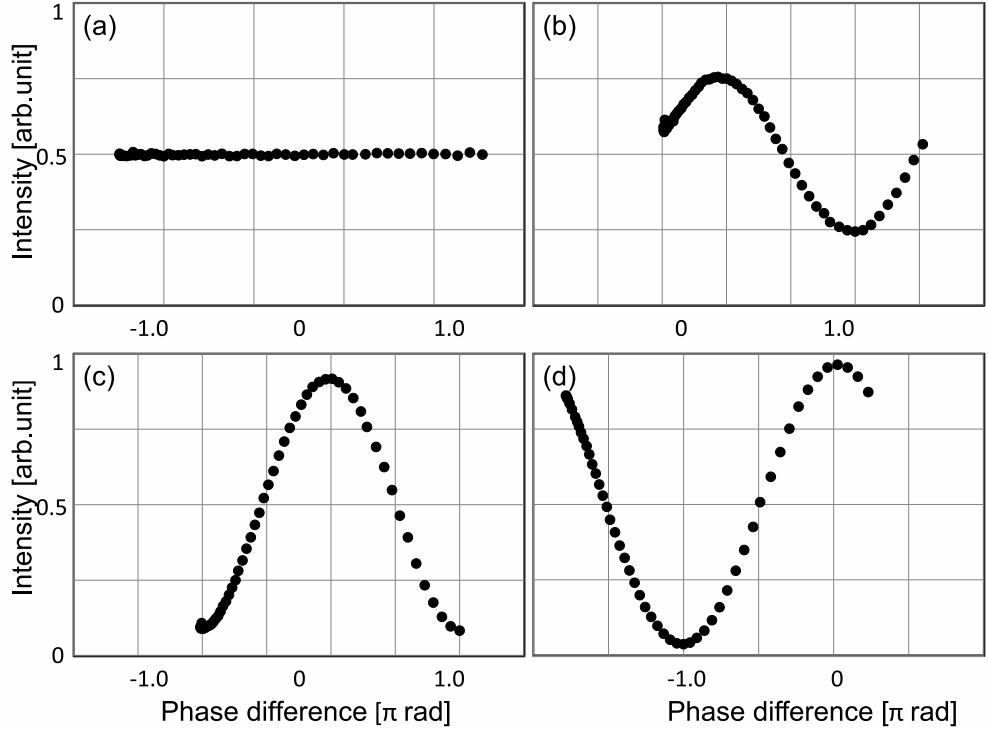}
\caption{
Interference fringes for several values of the normalized minimum excitation:
 (a) $\Lambda=-1.6$,  (b) $\Lambda=0.074$, (c) $\Lambda=0.49$,
  and (d) $\Lambda=2.6$. }
\label{fig:fringe}
\end{figure*}

The observed interference fringes are shown in Fig. \ref{fig:fringe}.
Clear interference fringe was observed for a large excitation ($\Lambda=2.6$)
 with the visibility close to unity ($\Theta=0.93$,)
  while no clear interference fringe was observed
   for a small excitation ($\Lambda=-1.6$.)
The results indicate that the phases of the pulses for a gain-switched LD
 are still random even at 10 GHz pulse frequency
  for a small minimum drive current.
In Fig. \ref{fig:fringe}, we plotted the normalized values of the output signal
 to set the averaged value to 0.5;
  $(\text{Intensity}) = (\text{Observed power}) /(\text{Averaged Power}) $.
Due to the limitation of the device, the range of phase modulation was 
only slightly larger than $2 \pi$. 
The phase difference of the interferometer was stable enough for the short time
 to obtain an interference fringe. 
It was not stable for days, so that the origin of the phase difference varied
 as shown in Fig. \ref{fig:fringe}(a)-(d.)  

\section{Discussion}
\label{sec:discussion}
We consider the origins of errors to examine the accuracy of the results
  obtained in the experiment. 
 First, the imperfections in the interferometer, such as fluctuation of
  path length, imbalanced branch ratio of the beam splitters,
  polarization rotation, and depolarization will reduce the visibility. 
 In fact, we obtained the visibility of only 0.95 with continuous wave (CW)
  light emitted from the LD excited solely by the DC current of 50 mA,
   which was far above the threshold. 
 The LD linewidth implies that the phase of the CW light should be
  well conserved in the time scale of 100 ps. 
 Therefore, we should consider the obtained visibility was affected mainly
  by the imperfections in the interferometer.
 Assuming the imperfections are the same throughout the experiment,
  we should correct the visibility by multiplying 1.05. 
  
\begin{figure*}
\centering
\includegraphics[width=1.0\linewidth]{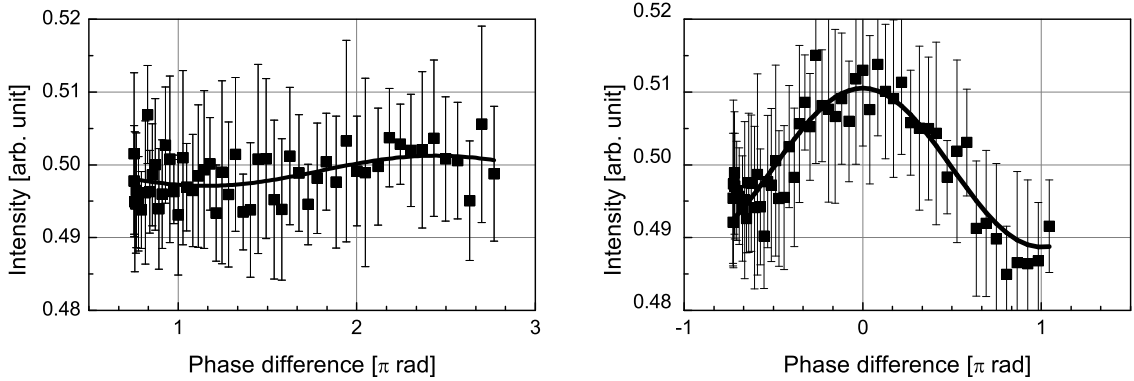}
\caption{Magnified view of the interference fringe.
 (a) $\Lambda=-1.6$ and (b) $ \Lambda=-1.2$. 
Solid lines denote the fitted curve.
 The fitted values of the visibility were 0.004 and 0.02. }
\label{fig:noise}
\end{figure*}

 Second, the system noise affects the visibility estimation. 
 In the present  experiment, we recorded the observed maximum
  and minimum values, which included noise. 
 Thus,  the estimated visibility should have been overestimated.
 This overestimation causes no harm,
  from the conservative points of view for the security certification.
 However, it  is undesirable for the practical use, because we may lose
  some amount of final key by unnecessary privacy amplification.
 The effect of the noise emerges significantly for small visibilities.
 To obtain  better estimation, we examined the results showing low visibilities
  by magnifying the scale of intensity, as shown in Fig. \ref{fig:noise}.
 The error bars originated mainly from the noise of the sampling oscilloscope. 
 The observed signal-to-noise ratio was about 17 dB,
  where the average intensity was normalized to 0.5.
 The r.m.s. value of the noise suggests that it may be hard to measure
  the visibility less than 0.02.     
 Nevertheless, a periodic dependence on the phase difference is seen
  in  Fig. \ref{fig:noise} (b). 
 By taking the center values denoted by squares in  Fig. \ref{fig:noise},
  we could fit the interference fringe with 
 \begin{equation}
  I(\varphi)=A \left(1+\Theta \cos(\varphi+\varphi_0) \right).
 \end{equation}    
The result of the fitting is depicted as a thick solid line
 in Fig. \ref{fig:noise}. 
For excitation of $\Lambda=-1.6$ (a),  the best fitting value for the visibility
 was 0.004, which reflected very weak periodic dependence
  on the phase difference. 
The fitted visibility was much less the one estimated from the maximum
 and minimum intensities,  0.014.  
For (b), $ \Lambda=-1.2$, the fitted value of the visibility was 0.022,
 while the estimated one was 0.030. 
The fit was done well, as we consider the 95 \% confidence interval of
 the fitting value [0.019, 0.025]. 
Discrepancy between the fitted and estimated visibility decreases
 as the visibility increases. 
On the basis of the above, we conclude that the present experimental set up
 can detect the visibility down to 0.02. 
The effect of the noise should be reduced by using low noise front-end
 and by increasing number of accumulation to obtain lower measurement limit
  of the visibility. 

\begin{figure}
 \centering
 \includegraphics[width=0.8\linewidth]{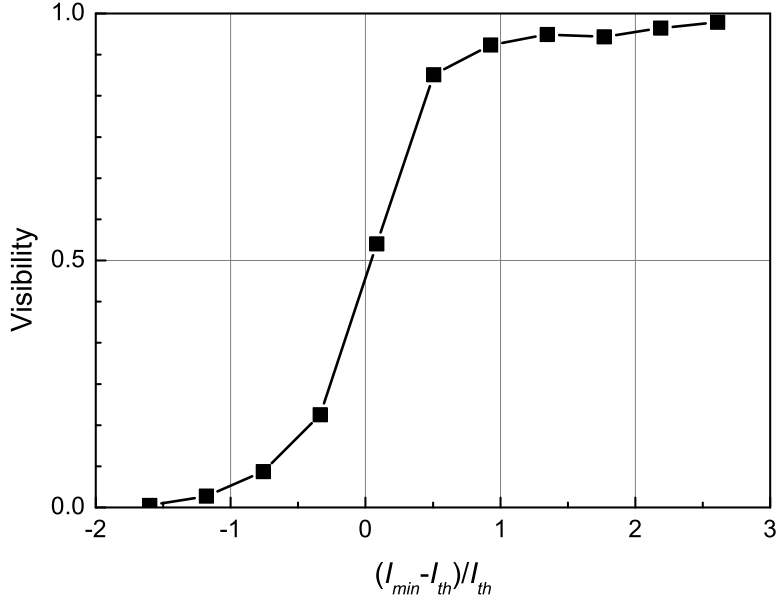}
 \caption{Visibility of the interference as a function of
  the normalized minimum excitation $\Lambda$. 
 For $\Lambda >0$, the LD was always turned on. 
 The LD was reversely biased at the bottom of the pulse, when $\Lambda <-1$.}
 \label{fig:result}
 \end{figure}

We applied the corrections discussed above. 
The results are summarized in Fig. \ref{fig:result},
  where the visibility is plotted as
   a function of the normalized minimum excitation.
Figure \ref{fig:result} shows the visibility increases
 as the minimum excitation.
It raises steeply around $\Lambda=0$, where the LD was always turned on.  
  The interference fringe almost disappeared when $\Lambda < -1$, i.e.,
   the LD was reversely biased at the bottom. 
In particular, for $\Lambda=-1.6$ ($I_{DC}=$30.95 mA,) the visibility was
 fitted to 0.004,  which satisfied the strictest criterion given
  in sec. \ref{sec:impact}.
Though the fitted value may not be accurate as described above,
 the visibility satisfied the target values 0.015 for the imbalanced coin
  at $\mu=0.01$, and 0.044 for the decoy state discrimination. 
It should be noted that the interference  fringe was observed
 even when the LD was turned off during the pulse interval. 
 When the minimum drive current was set in the range  $-1 < \Lambda < 0$,
  the light source can be no longer regarded as a phase randomized
   in terms of the imbalance of the quantum coin.
For example, the visibility reached 0.08 for $\Lambda=-0.76$.
If we care only about the laser waveform (as is common in most applications,)
 we may set the bias to the value where the minimum drive current is
  close to the LD threshold, because it yields the best waveform as seen
   in Fig. \ref{fig:lightpulse} (b) and (c). 
Unfortunately, the phases of the pulses are correlated
 under this operating condition. 
The observed visibility was 0.534 for the case (c), where $\Lambda=0.074$.  
The corresponding standard deviation of the phase distribution is
 about $\sigma =1.12$. 
We need to scarify more bits to guarantee the security of final key
 in the privacy amplification under this operating condition. 
 
In the following, we consider the dependence of the phase correlation
 on the minimum excitation in terms of effective photon life time. 
As described before, if the photons survive during the pulse interval,
 the phase may correlate with the previous pulses. 
Typical photon life time $\tau_{ph}$ of a LD cavity is several picoseconds.  
The effective photon lifetime can be increased by stimulated emission,
 even when the excitation is insufficient for lasing.
Photon density $S$ in the cavity will decay approximately as
\begin{equation}
 \frac{dS}{dt}=\left(\varGamma g(n)-\frac{1}{\tau_{ph}} \right)S+n_{sp},
 \label{photon_lifetime}
\end{equation}
where $\varGamma g(n)$ denotes the modal gain for the lasing mode
 at the carrier density $n$. 
The term $n_{sp}$ represents the contribution of the spontaneous emission
 to the lasing mode.
The photon field is governed by the spontaneous emission,
 when the photon density decreases to satisfy $S\le n_{sp}$.
Then, the phase of the light field become random. 
 
The details of the dynamics is described with involved nonlinear
 coupled equations on photon density and carrier density.
Roughly speaking, though, the photon field loses the phase information
 after the effective photon life time given by 
$-(\Gamma g(n)-1/\tau_{ph})^{-1}$, as seen in Eq. (\ref{photon_lifetime}).
Since $(\Gamma g(n)-1/\tau_{ph})\tau_{ph}$ equals approximately to
 $(I-I_{th})/I_{th}$, the effective photon lifetime scales
  with the inverse of the normalized excitation.
When the normalized minimum excitation $\Lambda$ exceeds zero,
 the effective photon life time becomes infinite to negative. 
Then, the photons of previous pulses remain to contribute the phase correlation.
Even when $\Lambda$ is less than zero, the photons may survive
 during the interval and contribute to the next lasing.
For example, at  $\Lambda = -0.33$,
 the effective photon the time life is about three times as large as
  the cavity life time at the bottom of the pulse.
If we take the cavity life time as $\tau_{ph}= 3$ ps,
 the effective photon lifetime increases to 10 ps.
This value is comparable to the turn-off duration  of 13 ps
calculated  from Eq.(\ref{current}). 
Therefore, a non-negligible number of  photons are supposed to remain
 under this condition. 
In fact, the observed visibility was 0.188, indicating some phase correlation.
For small excitation satisfying $\Lambda<-1$, the LD is reversely biased,
 and the effective photon life time should be equal to the cavity life time
  at least in the bottom of excitation. 
The calculated turn-off duration is as long as 30 ps for $\Lambda=-1.6$. 
Under this condition, photons should have disappeared during the pulse interval,
 and the lasing phase became random, as was observed in the experiment.

We see that observed dependence of the visibility on the excitation
 can be explained with the relation between the effective photon life time
  and the turn-off duration.
A guide of the operating condition can be summarized that
 the effective photon lifetime should be less than the turn-off duration.
As described above, this condition is satisfied with $\Lambda<-1$.
  
\section{Conclusion}
In BB84 protocol using an attenuated laser source,
 the secure key generation rate is lowered
  if the source emits non-phase randomized optical pulses.
We evaluated the effect of the phase correlation
 in terms of the imbalance of the quantum coin and the discrimination of
  the decoy  from the signal pulses, for the partially coherent states. 
We obtained criteria for the source to be regarded as phase randomized. 
The target values for the visibilities were 0.006 and 0.015
 in terms of the imbalance of the quantum coin at $\mu=0.09$ and at $\mu=0.01$,
 and 0.044 in terms of discrimination of the  decoy pulses ($\nu=0.1$)
  from the signal ($\mu=0.5$.)
We constructed a  phase correlation test system to measure visibility
 of the interference fringe between the adjacent pulses
  using an asymmetric Mach-Zehnder interferometer.
It enables to evaluate the phase correlation
 between the laser pulses experimentally at a high clock frequency of 10 GHz.
We found that the phase correlation of the pulses from a LD depends
 on the operating condition.   
The target values were satisfied with the gain-switched LD.
The operating condition is that the LD should be reversely biased
 at the bottom of the pulses, and the turn-off duration should be longer
  than the effective photon lifetime.
The results indicate that QKD system clock can be increased to 10-GHz
 as far as the security issue on the laser light source is concerned.
 In practice, a number of technical problems remain.
Among the remaining issues, the most important ones would be
 high speed photon detection and  post-processing.

\begin{acknowledgments}
The authors would like to thank Dr. Kiyoshi Tamaki and Dr. Yoshihiro Nambu
 for helpful discussions,
  and Mr. Takahisa Seki for his assistance in the experiment.
This work has been conducted under the commissioned research
 of National Institute of Information and Communication Technology (NICT,)
   "Secure photonic network technology." 
\end{acknowledgments}

\appendix
\section{Optimal discrimination of two density matrices}
\begin{figure}
\centering
\includegraphics[width=0.8\linewidth]{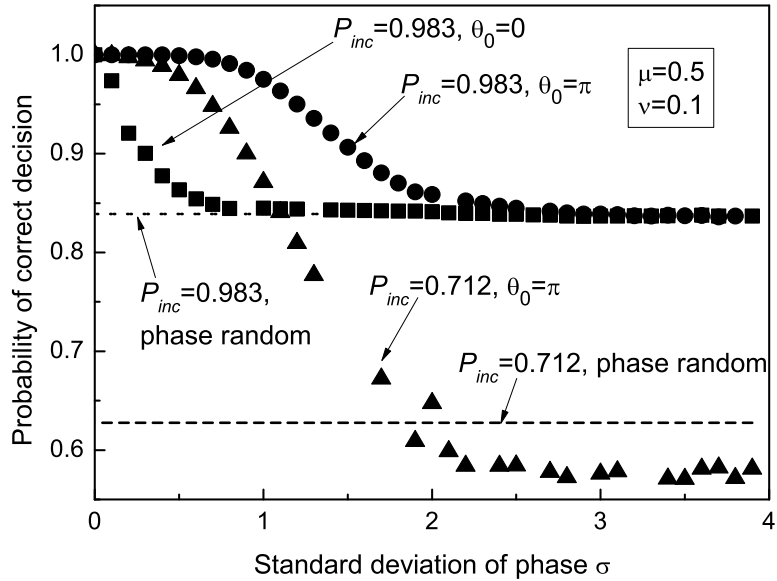}
\caption{Probability of correct decision
 as a function of the standard deviation of phase. 
State discrimination was done for the partially phase randomized states
 of average photon number 0.5 and 0.1.
Squares represent $P_{inc}=0.983,\: \theta_0=0$, circles  $P_{inc}=0.983,\:
 \theta_0=\pi$, and triangles  $P_{inc}=0.712,\: \theta_0=\pi$.
Dotted line and dashed line refer to the probability of correct decision
 on the phase randomized states under the conditions of  $P_{inc}=0.983$
  and $P_{inc}=0.712$, respectively. }
\label{fig:pcorrect}
\end{figure}

 We here derive the optimum discrimination of the decoy from the signal
  in partially coherent states, which were given in Eq. (\ref{eq:decoy}).
 We optimized the POVM for given values of $P_{inc}$ to obtain the highest $P_C$
  in discriminating the two mixed states. 
 All the density matrices here are real and symmetric.   
 The POVMs $\Pi_i,\: i=0,1,2$ satisfy
 \begin{equation}
  \Pi_0+\Pi_1+\Pi_2=1,
  \label{povmcond1}
 \end{equation}
where $\Pi_1$ and $\Pi_2$ correspond to the conclusive decision
 that the state is $\rho_1$ and $\rho_2$, respectively,
  while $\Pi_0$ represents the inconclusive results.
The probability of inconclusive results is given by
\begin{equation}
P_{inc}= \mathrm{Tr}\left(\rho \Pi_0 \right) 
=1- \mathrm{Tr}\left(\rho (\Pi_1+\Pi_2) \right),
\label{pinc}
\end{equation}
and the probability to yield a correct decision is
\begin{align}
 P_C &= p \mathrm{Tr} \left(\rho_1 \Pi_1 \right)+
       (1-p)  \mathrm{Tr} \left(\rho_2 \Pi_2 \right) \nonumber \\
     &= 1-P_{inc} - p \mathrm{Tr} \left(\rho_1 \Pi_2 \right)
       -(1-p) \mathrm{Tr} \left(\rho_2 \Pi_1 \right),
   \label{pcorrect} 
\end{align}
where $\rho$ is defined with  the probability of $\rho_1$'s occurrence $p$ by
\begin{equation}
\rho=p \rho_1+(1-p) \rho_2.
\end{equation}
We applied the  iteration method developed by Fiur\'a\v{s}ek and
 Je\v{z}ek \cite{FJ2003odo} to maximize $P_C$ (\ref{pcorrect})
  under the constraints (\ref{povmcond1}) and (\ref{pinc})
   for a given value of $P_{inc}$ using Lagrange multipliers. 
The iteration was performed with symmetrized equations to keep
 the POVMs Hermite and positive semi-definite. 
We set average photon numbers for the signal and decoy state 
 $\mu= 0.5$ and $\nu=0.1$, respectively, 
and assumed the signal and decoy state appear
 with the same probability ($p=1/2$) for simplicity. 
The optimization was done for two values of the probability of
 the inconclusive results: $P_{inc}=0.983$,
  which refers to the probability of the inconclusive results
   in the USD measurement to weak coherent states 
   of $\theta_0=0$, and $P_{inc}=0.712$ of $\theta_0=\pi$.
No USD measurement for  $\theta_0=0$ exists to satisfy  $P_{inc}=0.712$.
Figure \ref{fig:pcorrect} shows the probability of correct decision
 as a function of the standard deviation of phase.
Successful USD measurement ($P_C=1$) is achieved
 for the coherent states ($\sigma=0$.) 
The probabilities of correct decision decrease as the standard deviation
 of phase increasing. 
The probabilities converge to the values for completely phase randomized states,
 when $\sigma$ exceeds about 2.5, which corresponds to the visibility of 0.044.
    
We observed that the numerical optimization under the condition of
  $P_{inc}=0.712$ sometimes failed, in particular for  large $\sigma$. 
It also returned the  probability of correct decision lower than that
 for the phase randomized states. 
Small errors in matrix operations may result in such suboptimal results. 
Nevertheless, we found the optimized POVMs for large $\sigma$ states were
 almost the same as those for the phase randomized states. 
The optimized POVM operators were almost diagonal in photon-number state basis. 
The outcome $\rho_1$ was obtained when the observed photon number was
 larger than a threshold, while the inconclusive result was obtained
  when it was smaller than the threshold. 
The given probability of the inconclusive results determined
 the threshold photon number.     

\bibliography{phasecorrelation}
 
\end{document}